# Neural-Optic Co-Designed Polarization-Multiplexed Metalens for Compact Computational Spectral Imaging

Qiangbo Zhang[1,†], Peicheng Lin[2,†], Chang Wang[1], Yang Zhang[1], Zeqing Yu[1], Xinyu Liu[1], Ting Xu[2,*] and Zhenrong Zheng[1,*]

[1]College of Optical Science and Engineering, Zhejiang University, Hangzhou 310027, China

[2]National Laboratory of Solid-State Microstructures, Collaborative Innovation Center of Advanced Microstructures, College of Engineering and Applied Sciences, Nanjing University, Nanjing, China

[†] These authors contributed equally: Qiangbo Zhang, Peicheng Lin

* Correspondence: xuting@nju.edu.cn (T. Xu), zzr@zju.edu.cn (Z. Zheng)

As the realm of spectral imaging applications extends its reach into the domains of mobile technology and augmented reality, the demands for compact yet high-fidelity systems become increasingly pronounced. Conventional methodologies, exemplified by coded aperture snapshot spectral imaging systems, are significantly limited by their cumbersome physical dimensions and form factors. To address this inherent challenge, diffractive optical elements (DOEs) have been repeatedly employed as a means to mitigate issues related to the bulky nature of these systems. Nonetheless, it's essential to note that the capabilities of DOEs primarily revolve around the modulation of the phase of light. Here, we introduce an end-to-end computational spectral imaging framework based on a polarization-multiplexed metalens. A distinguishing feature of this approach lies in its capacity to simultaneously modulate orthogonal polarization channels. When harnessed in conjunction with a neural network, it facilitates the attainment of high-fidelity spectral reconstruction. Importantly, the framework is intrinsically fully differentiable, a feature that permits the joint optimization of both the metalens structure and the parameters governing the neural network. The experimental results presented herein validate the exceptional spatial-spectral reconstruction performance, underscoring the efficacy of this system in practical, real-world scenarios. This innovative approach transcends the traditional boundaries separating hardware and software in the realm of computational imaging and holds the promise of substantially propelling the miniaturization of spectral imaging systems.

**Introduction**

In the contemporary age of extensive data generation, there is a growing need for acquiring detailed visual information. Conventional RGB cameras, which provide two-dimensional data, have their limitations in offering a comprehensive perspective. Spectral imaging, on the other hand, captures three-dimensional data cubes that encompass both spatial and spectral data, providing richer data information[1–4]. Simultaneously, the proliferation of mobile devices and the increasing prevalence of augmented reality (AR) and virtual reality (VR) technologies have amplified the requirement for compact imaging solutions[5–7]. Consequently, the research focus has shifted towards the development of miniaturized spectral imaging systems to meet this demand[8–11].

To facilitate the real-time capture of dynamic three-dimensional spectral data cubes, coded aperture snapshot spectral imaging (CASSI) systems have gained prominence[12,13]. However, the bulkiness of these systems, attributed to the integration of multiple refractive optical elements, significantly hampers their portability. Efforts to address this challenge have led to the incorporation of diffractive optical elements (DOEs) into snapshot spectral imaging systems, resulting in more compact designs[8–10]. Despite these advancements, DOEs are restricted to phase modulation of light, thus limiting their capabilities for achieving high-fidelity imaging. In contrast, the emergence of a new generation of flat optical elements, known as metasurfaces, offer the potential to modulate not only the phase but also the amplitude and polarization of light[14–20], thus paving the way for the development of miniaturized spectral imaging systems.

Metalenses have found extensive applications across various computational imaging techniques, including full-color achromatic imaging[21–23], depth estimation[24–29], extended depth-of-field[23,30,31] and spectral imaging[11,32,33]. However, in most existing works, the optimization of the front-end metalens and the back-end recovery algorithm are treated as separate endeavors, neglecting the core principle of computational imaging, which involves the joint optimization of both hardware and software. Although there are few works tended to use end-to-end differentiable framework to design metalens for realizing computational imaging[21,34,35], a comprehensive investigation centered on a fully differentiable computational spectral imaging system based solely on a single metalens remains a largely uncharted area within the realm of research.

In this work, we introduce a comprehensive end-to-end computational spectral imaging system based on a polarization-multiplexed metalens. Utilizing a single metalens as the sole optical imaging component, we combine it with a specialized polarization camera to independently modulate the orthogonal polarization channels. This encoding strategy is supported by a neural network-based decoding framework, contributing to high-quality spectral reconstruction. A unique feature of this system is its full differentiability, covering the parameterization of individual metalens units, the compressive imaging stage, and the neural network-driven spectral image recovery. This enables a synergistic optimization of both the front-end metalens and the back-end neural network, maximizing the coding and decoding capabilities of each and surpassing the performance of systems optimized in isolation. We validate the theoretical contributions by constructing a prototype camera incorporating a polarization-multiplexed metalens. In real-world tests, our system achieves remarkable spectral recovery while preserving a wealth of details. We believe that this innovative computational spectral approach will advance the development of spectral imaging and promote the miniaturization of such systems.

**Results and discussions**

**End-to-end computational spectral imaging framework**

We present a computational spectral imaging system based on a polarization-multiplexed metalens, as illustrated in Fig. 1a. By employing a polarization-sensitive metalens, the scene's spectral components are projected onto a specialized polarization camera. Images captured in orthogonal $x$ and $y$ polarization states are then processed via a dual-input ResUNet architecture, facilitating precise reconstruction of the spectral image. Figure 1b outlines the end-to-end computational spectral imaging framework. The process begins with the per-pixel metalens unit structure, which is linked to its complex amplitude response through a pretrained multi-layer perceptron (MLP). It establishes the complex amplitude distribution at the metalens' exit surface. These unit structures of the metalens exhibit polarization dependence, resulting in distinct complex amplitude distributions under $x$ and $y$ polarized incidences. By applying the Fresnel diffraction principle, we calculate dual series of polarization-sensitive point spread functions (PSFs). These PSFs are then convolved with the spectral image, considered as the ground truth, to generate polarization-specific optical responses. Subsequently, the convolved images are compressed into noisy RGB representations, leveraging the camera's inherent spectral response to simulate the real-world, blurry RGB images in both $x$ and $y$ polarization channels. These blurred images are then transformed back into their spectral form using a specialized dual-input neural network. The reconstructed spectral image is compared against the ground truth to derive a loss function, which is backpropagated to synergistically optimize both the metalens structure and the network parameters.

**Metalens unit structure response using MLP**

The essence of end-to-end computational imaging lies in ensuring differentiability throughout each process. In this work, we employ a straightforward MLP to encapsulate the transformation from the metalens unit structure to the complex amplitude response, guaranteeing complete differentiability[34,35]. Specifically, we make use of the polarization-dependent unit structure illustrated in Fig. 2a, utilizing silicon dioxide ($SiO_2$) as the substrate and silicon nitride ($Si_3N_4$) as the nanofin material.

The initial phase of our methodology involves the application of the full-wave finite-difference time-domain (FDTD) method to construct a comprehensive database. This endeavor entails a systematic exploration of the nanofin's geometric attributes under diverse wavelengths, thereby establishing associations between the unit structure and its respective responses under various polarization states. This database serves as the training data for MLP fitting. The input parameters include the nanopillar's length $L$ and width $W$, and the incident wavelength $\lambda$, while the output parameters comprise transmittance $T$ and phase shifts $\varphi$ under $x$ and $y$ polarization channels, expressed as follows:

$$\{L, W, \lambda\} \rightarrow \{T_x, \sin\varphi_x, \cos\varphi_x, T_y, \sin\varphi_y, \cos\varphi_y\}. \qquad (1)$$

The subscripts represent different polarization channels. We employ trigonometric functions to denote phase delays, mitigating phase convexity's influence on MLP fitting. Our MLP features two hidden layers, each equipped with 512 neurons. The activation layer employs a Leaky ReLU function, and the last layer uses a Tanh function to restrict the trigonometric representation of phase delay within an interval from -1 to 1. Transmittance values are clamped to remain within the range of 0 to 1. The loss function employed is mean squared error (MSE) loss.

Figure 2b presents a comparative analysis between MLP-fitted responses and results obtained through FDTD simulations. This comparison demonstrates the successful modeling optical responses for nanofins with varying dimensions under various wavelengths and polarization states by our MLP. As a result, the differentiable MLP layer stands as the foundational step in our end-to-end computational spectral imaging framework.

**Spectral image formation model.**

After establishing a differentiable model from the metalens structure to the response, the subsequent phase involves simulating the light propagation process from the metalens to the sensor. In pursuit of enhanced computational efficiency, we adopt a rotationally symmetric distribution for the metalens structure[36,37], as visually depicted in Fig. 2c. This approach allows for a simplification of the PSF into a one-dimensional calculation. The PSF, contingent upon both varying wavelengths and distinct polarization states, is mathematically represented as:

$$\text{PSF}_\lambda^{pol} = \int F(P^{pol}, \rho, \lambda, f) d\rho, \tag{2}$$

where the superscript *pol* discriminates among unique polarization channels, specifically 0° and 90°. The variable $\rho$ stands as the radial coordinate within the PSF, while $\lambda$ signifies the wavelength of the incident light. The focal length of the metalens is represented by *f*, and the function $P^{pol}$ represents the complex amplitude responses of the metalens across radial directions under distinct polarization states. Further details are given in Supplementary Information.

Following the acquisition of PSFs, we proceed to simulate the imaging result of a spectral image (ground truth) captured by a polarization camera. The spectral image is convolved with the dual-channel PSFs and compressed to RGB channels through the spectral response of the polarized camera. Accounting for the inherent influence of noise, the final RGB image, denoted as $J_{RGB}$, captured by the polarized camera can be expressed as:

$$J_{RGB}^{pol} = \int \left( I_\lambda * PSF_\lambda^{pol} \right) \cdot S_{\lambda,rgb}^{pol} d\lambda + n, \tag{3}$$

where $I_\lambda$ denotes the original spectral image, $S_{\lambda,rgb}$ stands for the spectral sensitivity function of the sensor, and the symbol $*$ represents the convolution operator.

**Spectral image recovery neural network.**

In light of the exemplary decoding capabilities exhibited by the U-Net framework in computational imaging[38], we adopt a variant of this model, termed ResUNet[39], as the foundational architecture for neural network-based spectral image restoration in this investigation. Nevertheless, our approach diverges from the conventional ResUNet framework in a crucial manner. Specifically, by leveraging a polarization-sensitive metalens and a polarization camera, we capture RGB images under both 0° and 90° linear polarization channels, thereby acquiring supplementary encoding information. To fully harness this enriched encoding data, our decoding process employs a Dual-Input ResUNet (DI-ResUNet), as visually represented in Fig. 2d.

The DI-ResUNet architecture comprises two identical contracting paths and a singular expansive path. Each contracting path commences with an image that undergoes a double residual convolutional block, facilitating the extraction of 32 feature dimensions. The image subsequently progresses into a layer featuring a down-sampling block and a double residual convolutional block. This sequence of operations is iteratively performed four times within each contracting path. Subsequently, the features extracted from the two contracting paths are amalgamated to serve as the starting point for the expansive path. The expansive path mirrors the operations of the contracting path, commencing with up-sampling and followed by a double residual convolutional block. In order to preserve fine image details, the corresponding blocks from the two contracting paths are concatenated. Finally, a residual convolutional block is introduced at the terminus of the expansive path to yield the restored spectral image $\hat{I}_\lambda$ and the loss function is defined as $Loss = \left\| I_\lambda - \hat{I}_\lambda \right\|_2^2$. Further details are available in Supplementary Information.

**Implementation example.**

We implement our proposed end-to-end computational spectral imaging framework based on a metalens with PyTorch, successfully achieving spectral imaging within the range of 420-660 nm at an interval of 10 nm. Specifically, we design a polarization-dependent metalens with a diameter of 2.8 mm. The metalens is positioned 20mm in front of the polarization camera sensor. Our optimization process initiates at the unit structure of the metalens, necessitating an additional initialization step. This process focuses 530 nm linearly polarized light at 0° and 90° onto points located 20 mm and 23 mm after the metalens, respectively. We use hyperspectral dataset publicly available from ICVL[40], Harvard[41], and KAIST[42] datasets. For testing purposes, 20 images are randomly selected from the ICVL dataset and 10 images from the KAIST dataset, while the remaining images are used for training. Further details are available in Supplementary Information.

**Assessment in simulation**

Our specially designed polarization-dependent metalens enables the capture of two distinct sets of PSFs under $x$ and $y$ linear polarizations. Initially, our metalens focuses 530 nm $x$-polarized light onto the polarization sensor, providing detailed image information. Simultaneously, 530 nm $y$-polarized light is directed to be focused 3 mm behind the polarization sensor, contributing supplementary spectral data. As a result, the size of the PSFs under $x$-polarization is generally smaller in comparison to those under $y$-polarization across different wavelengths.

Within the framework of our end-to-end computational spectral imaging, the unit structure of the metalens and the parameters of the DI-ResUNet are jointly optimized. Upon the completion of this joint optimization, the one-dimensional radial unit structure is cubically interpolated onto a two-dimensional plane. This process yields the recalculated PSFs under the two polarization channels, as depicted in Fig. 4e. Notably, these PSFs exhibit distinctions for the same wavelength under different polarization channels, as well as for different wavelengths under the same polarization channel. These variations among the PSFs play a pivotal role in uniquely encoding diverse spectral information from the scene, which in turn facilitates the subsequent decoding process by the back-end neural network.

The first two panels of Fig. 3a display the images captured by the polarization camera under *x* and *y* polarization channels, respectively. A noteworthy aspect of these images is the notable variance in color both among corresponding pixels and within distinct regions of each image. Specifically, the image under *x*-polarization renders more explicit color details for shorter wavelengths, while the image under *y*-polarization offers clearer color details for longer wavelengths. This observation is consistent with the results of our optimization process, which indicate that the smallest PSF size occurs at 530 nm under *x*-polarization and at 610 nm under *y*-polarization. The latter pair of panels in Fig. 3a depict the spectral images generated by the DI-ResUNet and the original spectral images (ground truth), respectively. Upon visual inspection, the two sets of images exhibit a remarkable degree of similarity.

To further underscore the effectiveness of our proposed computational imaging system, which is built upon the foundation of polarization multiplexing metalens and a comprehensive hardware-software optimization approach (PM+Joint), we carried out comparative experiments. The control groups included systems using polarization multiplexing metalens but optimizing only the back-end network (PM+Net), polarization-independent metalens with joint software-hardware optimization (PI+Joint), and polarization-independent metalens optimizing only the back-end network (PI+Net). The detailed description of the polarization-independent metalens design is available in Supplementary Information.

We utilize three quantitative image quality metrics to evaluate the performance of these methodologies: the average peak signal-to-noise ratio (PSNR), structural similarity (SSIM), and root mean square error (RMSE). The results of the spectral image recovery, conducted on both the ICVL and KAIST datasets, are presented in Table 1, which summarizes the numerical outcomes for our approach and the control groups. Notably, our method attains the highest PSNR and SSIM scores while registering the lowest RMSE values on both datasets. For a more visually comprehensible representation of our findings, two representative sets of restored images are selected from the test dataset, as shown in Fig. 3b. It is obvious that our proposed method not only excels in image quality metrics but also exhibits sharper edges and clearer details, surpassing the results achieved by other methods. Furthermore, Fig. 3c showcases the spectra recovered through various methods for two randomly selected points within the chosen image. The spectra reconstructed using the proposed method closely align with the ground truth. Overall, our approach distinctly outperforms alternative methods in terms of both spatial image quality and the faithfulness of spectral curve reconstruction, thus reinforcing the findings derived from the numerical comparisons.

**Experimental implementation**

After completing the end-to-end optimization, we proceed to fabricate a polarization-sensitive metalens in strict accordance with our designed specifications. The fabricated metalens is depicted in Fig. 4a. To facilitate this integration, we engineer a mechanical mounting casing capable of accommodating both the metalens and a broadband filter covering the spectral range from 420 to 660 nm. This assembly is seamlessly integrated into a polarization camera (Sony IMX250MYR). A visual representation of the resulting experimental prototype is provided in Fig. 4b. The optical microscopy image of the fabricated metalens, as shown in Fig. 4c, demonstrates rotational symmetry that is consistent with our design. Figure 4d presents the scanning electron microscope (SEM) images of the $Si_3N_4$ metalens. The top-view and perspective-view images distinctly reveal nanopillars with remarkable precision, aligning closely with our original design specifications. A comprehensive description of the fabrication process is given in the Methods section.

Recognizing the inherent potential for fabrication discrepancies and inaccuracies during the assembly stage, we implement a crucial PSF calibration step to accurately measure the PSF across different wavelengths under both $x$ and $y$ polarization states. In this process, white light from an LED source passes through a 50 μm pinhole and is subsequently collimated using a lens. The collimated light is then directed through a narrowband filter designed for a specified wavelength before reaching the sensor. Detailed schematics of the PSF capture setup is available in the Supplementary Information (Figure S8). The PSFs captured by the polarization sensor at various wavelengths are showcased in Fig. 4f, revealing minor deviations when compared to the simulated optimized PSFs. Leveraging the experimentally obtained PSFs, we incorporate these measured PSFs as fixed parameters within our optimization framework. This approach facilitates the fine-tuning of the back-end DI-ResUNet to mitigate discrepancies arising from metalens fabrication and mechanical assembly, thereby more accurately mirroring real-world imaging scenarios and subsequent spectral recovery.

With our experimental setup, we conduct imaging sessions encompassing both indoor and outdoor scenes, as depicted in Fig. 5a. The first two columns of Fig. 5a showcase images directly captured by our apparatus under $x$ and $y$ polarization states. In the first indoor scene featuring a color card, distinct color details in the green spectrum are conspicuously present under the $x$-polarized channel, while the $y$-polarized channel accentuates clearer boundaries for longer wavelengths, such as yellow and red. This observation is in alignment with the simulation results presented in Fig. 3a. By processing images acquired under varying polarization states through our optimized DI-ResUNet network, we achieve high-quality spectral reconstructions, as evidenced in the third column of Fig. 5a. The restored images exhibit visually pleasing outcomes. The spectral curves extracted from marked points in Fig. 5a are presented in Fig. 5b. These curves closely align with the spectra measured using commercial spectrometers, proving the effectiveness of our spectral reconstruction technique.

**Discussion**

In our proposed computational spectral imaging based on polarization multiplexing metalens, the fundamental assumption is that the light reflected by the object under investigation is unpolarized. Therefore, if the illuminating light is initially polarized or the material properties of the object introduce alterations to the polarization state of the incident light, our system may no longer be applicable or may require adjustments to account for these polarization effects. Furthermore, while the sensor features four linear polarization channels (0°, 45°, 90° and 135°), we have only exploited the 0° and 90° channels in our current implementation. Although the information content across all four channels may exhibit redundancy, the incorporation of the 45° and 135° polarization states into a more advanced network architecture could potentially lead to enhanced performance in spectral image reconstruction. This prospective avenue for improvement will be a central focus of our future research endeavors.

**Conclusion**

In summary, we have introduced an end-to-end computational spectral imaging system that leverages a polarization-multiplexing metalens and have validated this concept through the creation of a physical device. By utilizing a polarization-dependent metalens, we can concurrently modulate light in both the *x* and *y* polarization channels, in conjunction with a polarization camera to capture additional encoded information. Through the implementation of an end-to-end differentiable framework, we jointly optimize the structural parameters of the front-end metalens and the back-end neural network, resulting in the achievement of high-fidelity computational spectral imaging. In comparison to polarization-independent metalenses, our designed metalens offers increased degrees of freedom for encoding spectral information, consequently realizing superior spectral image recovery quality. Our proposed end-to-end computational spectral imaging system based on a polarization-multiplexing metalens establishes a further connection between metalens and computational imaging, and is expected to contribute to the ongoing miniaturization and advancement of high-performance spectral imaging technologies.

**Methods**

Metalens fabrication

The proposed metalens is fabricated through a standard top-down nanofabrication process including e-beam lithography (EBL) and reactive ion etching (RIE) technique. First, a layer of 800-nm-thick $Si_3N_4$ film is coated on a 500-μm-thick fused silica substrate by plasma enhanced chemical vapor deposition (Oxford, PlasmaPro 100 PECVD) at 300 ℃. Then, the surface of $Si_3N_4$ film is prepared for oxygen plasma cleaning and hexamethyldisilazane (HMDS) vapor-coating to improve adhesion. Afterwards, a layer of 200-nm-thick positive e-beam resist (ARP6200) is spin-coated on the $Si_3N_4$ film and baked on a hot plate for 3 minutes at 180 °C. To suppress the charge accumulation effect during the e-beam lithography process, an e-spacer layer (ARPC5090) is subsequently spin-coated on the resist. Hereafter, EBL (Elionix, ELS-F125) with an acceleration voltage of 125 kV and development are done in turn to define the designed pattern in the resist. After depositing a layer of 30-nm-thick aluminum on the structural resist by an e-beam evaporator (SKY, DZS500), a lift-off process in n-methyl-pyrrolidone (NMP) is carried out for 10 minutes at 80 ℃, leading to pattern reversion from the resist to the aluminum hard mask. So as to construct the $Si_3N_4$ nanopillars, inductively coupled plasma reactive ion etching (Tailong Electronics, ICP-

200) is implemented with a mixture of $C_4F_8$ and $SF_6$, where the ratio of $C_4F_8/SF_6$ is tuned to be 3. The bias RF power is 40 W, the ICP generator RF power is 500 W and the pressure is maintained at 16 mTorr. Finally, the metalens composed of $Si_3N_4$ nanopillars is obtained after removing the residual aluminum with the stripping solution (Sigma-Aldrich Etchant Type A).

**Reference**
1. Huang, L., Luo, R., Liu, X. & Hao, X. Spectral imaging with deep learning. *Light Sci. Appl.* **11**, 61 (2022).
2. Lu, G. & Fei, B. Medical hyperspectral imaging: a review. *J. Biomed. Opt.* **19**, 010901 (2014).
3. Li, Q. *et al.* Review of spectral imaging technology in biomedical engineering: achievements and challenges. *J. Biomed. Opt.* **18**, 100901 (2013).
4. Liang, H. Advances in multispectral and hyperspectral imaging for archaeology and art conservation. *Appl. Phys. A* **106**, 309–323 (2012).
5. Li, Y. *et al.* Ultracompact multifunctional metalens visor for augmented reality displays. *PhotoniX* **3**, 29 (2022).
6. Cheng, D. *et al.* Design and manufacture AR head-mounted displays: A review and outlook. *Light Adv. Manuf.* **2**, 336 (2021).
7. Kress, B. C. & Pace, M. Holographic optics in planar optical systems for next generation small form factor mixed reality headsets. *Light Adv. Manuf.* **3**, 1–31 (2022).
8. Jeon, D. S. *et al.* Compact snapshot hyperspectral imaging with diffracted rotation. *ACM Trans. Graph.* **38**, 1–13 (2019).
9. Baek, S.-H. *et al.* Single-shot Hyperspectral-Depth Imaging with Learned Diffractive Optics. in *2021 IEEE/CVF International Conference on Computer Vision (ICCV)* 2631–2640 (IEEE, 2021).
10. Li, L. *et al.* Quantization-aware Deep Optics for Diffractive Snapshot Hyperspectral Imaging. in *2022 IEEE/CVF Conference on Computer Vision and Pattern Recognition (CVPR)* 19748–19757 (IEEE, 2022).
11. Hua, X. *et al.* Ultra-compact snapshot spectral light-field imaging. *Nat. Commun.* **13**, 2732 (2022).
12. Gehm, M. E., John, R., Brady, D. J., Willett, R. M. & Schulz, T. J. Single-shot compressive spectral imaging with a dual-disperser architecture. *Opt. Express* **15**, 14013 (2007).
13. Wagadarikar, A., John, R., Willett, R. & Brady, D. Single disperser design for coded aperture snapshot spectral imaging. *Appl. Opt.* **47**, B44 (2008).
14. Khorasaninejad, M. *et al.* Metalenses at visible wavelengths: Diffraction-limited focusing and subwavelength resolution imaging. *Science* **352**, 1190–1194 (2016).
15. Khorasaninejad, M. & Capasso, F. Metalenses: Versatile multifunctional photonic components. *Science* **358**, eaam8100 (2017).
16. Banerji, S. *et al.* Imaging with flat optics: metalenses or diffractive lenses? *Optica* **6**, 805 (2019).

17. Engelberg, J. & Levy, U. The advantages of metalenses over diffractive lenses. *Nat. Commun.* **11**, 1991 (2020).
18. Rubin, N. A., Shi, Z. & Capasso, F. Polarization in diffractive optics and metasurfaces. *Adv. Opt. Photonics* **13**, 836–970 (2021).
19. Chen, W. T. & Capasso, F. Will flat optics appear in everyday life anytime soon? *Appl. Phys. Lett.* **118**, 100503 (2021).
20. Pan, M. *et al.* Dielectric metalens for miniaturized imaging systems: progress and challenges. *Light Sci. Appl.* **11**, 195 (2022).
21. Tseng, E. *et al.* Neural nano-optics for high-quality thin lens imaging. *Nat. Commun.* **12**, 6493 (2021).
22. Colburn, S., Zhan, A. & Majumdar, A. Metasurface optics for full-color computational imaging. *Sci. Adv.* **4**, eaar2114 (2018).
23. Huang, L., Whitehead, J., Colburn, S. & Majumdar, A. Design and analysis of extended depth of focus metalenses for achromatic computational imaging. *Photonics Res.* **8**, 1613 (2020).
24. Jin, C., Zhang, J. & Guo, C. Metasurface integrated with double-helix point spread function and metalens for three-dimensional imaging. *Nanophotonics* **8**, 451–458 (2019).
25. Jin, C. *et al.* Dielectric metasurfaces for distance measurements and three-dimensional imaging. *Adv. Photonics* **1**, 1 (2019).
26. Lin, R. J. *et al.* Achromatic metalens array for full-colour light-field imaging. *Nat. Nanotechnol.* **14**, 227–231 (2019).
27. Guo, Q. *et al.* Compact single-shot metalens depth sensors inspired by eyes of jumping spiders. *Proc. Natl. Acad. Sci.* **116**, 22959–22965 (2019).
28. Tan, S., Yang, F., Boominathan, V., Veeraraghavan, A. & Naik, G. V. 3D Imaging Using Extreme Dispersion in Optical Metasurfaces. *ACS Photonics* **8**, 1421–1429 (2021).
29. Jing, X. *et al.* Single-shot 3D imaging with point cloud projection based on metadevice. *Nat. Commun.* **13**, 7842 (2022).
30. Bayati, E. *et al.* Inverse Designed Metalenses with Extended Depth of Focus. *ACS Photonics* **7**, 873–878 (2020).
31. Fan, Q. *et al.* Trilobite-inspired neural nanophotonic light-field camera with extreme depth-of-field. *Nat. Commun.* **13**, 2130 (2022).
32. Yang, J. *et al.* Ultraspectral Imaging Based on Metasurfaces with Freeform Shaped Meta-Atoms. *Laser Photonics Rev.* **16**, 2100663 (2022).
33. Xiong, J. *et al.* Dynamic brain spectrum acquired by a real-time ultraspectral imaging chip with reconfigurable metasurfaces. *Optica* **9**, 461 (2022).
34. Hazineh, D. S. *et al.* D-Flat: A Differentiable Flat-Optics Framework for End-to-End Metasurface Visual Sensor Design. Preprint at http://arxiv.org/abs/2207.14780 (2022).
35. Hazineh, D., Lim, S. W. D., Guo, Q., Capasso, F. & Zickler, T. Polarization Multi-Image Synthesis with Birefringent Metasurfaces. in *2023 IEEE International Conference on Computational Photography (ICCP)* 1–12 (IEEE, 2023).
36. Dun, X. *et al.* Learned rotationally symmetric diffractive achromat for full-spectrum computational imaging. *Optica* **7**, 913 (2020).
37. Ikoma, H., Nguyen, C. M., Metzler, C. A., Peng, Y. & Wetzstein, G. Depth from Defocus with Learned Optics for Imaging and Occlusion-aware Depth Estimation. in *2021 IEEE International Conference on Computational Photography (ICCP)* 1–12 (IEEE, 2021).

38. Ronneberger, O., Fischer, P. & Brox, T. U-Net: Convolutional Networks for Biomedical Image Segmentation. in *Medical Image Computing and Computer-Assisted Intervention – MICCAI 2015* (eds. Navab, N., Hornegger, J., Wells, W. M. & Frangi, A. F.) vol. 9351 234–241 (Springer International Publishing, 2015).
39. Xiao, X., Lian, S., Luo, Z. & Li, S. Weighted Res-UNet for High-Quality Retina Vessel Segmentation. in *2018 9th International Conference on Information Technology in Medicine and Education (ITME)* 327–331 (IEEE, 2018).
40. Arad, B. & Ben-Shahar, O. Sparse Recovery of Hyperspectral Signal from Natural RGB Images. in *Computer Vision – ECCV 2016* (eds. Leibe, B., Matas, J., Sebe, N. & Welling, M.) vol. 9911 19–34 (Springer International Publishing, 2016).
41. Chakrabarti, A. & Zickler, T. Statistics of real-world hyperspectral images. in *CVPR 2011* 193–200 (IEEE, 2011).
42. Choi, I., Jeon, D. S., Nam, G., Gutierrez, D. & Kim, M. H. High-quality hyperspectral reconstruction using a spectral prior. *ACM Trans. Graph.* **36**, 1–13 (2017).

**Acknowledgements**

The work is supported by the National Key Research and Development Program of China (2022YFF0705500), National Natural Science Foundation of China (12274217), Natural Science Foundation of Jiangsu Province of China (BK20220068). The microfabrication center of the National Laboratory of Solid-State Microstructures of Nanjing University is acknowledged for their technique support. We thank Chenying Yang and Haiquan Hu for helpful discussions.

**Author contributions**

Q.Z., T.X. and Z.Z. conceived the idea. Q.Z., C.W., Z.Y. and X.L. performed the numerical simulations. Q.Z., P.L., C.W., Y.Z. and Z.Y. performed the experimental measurements. P.L. fabricated the sample. Q.Z., P.L., T.X. and Z.Z. contributed to the interpretation and analysis of results. T.X. and Z.Z. directed the project. All authors participated in paper preparation.

**Competing interests**

The authors declare no competing interests.

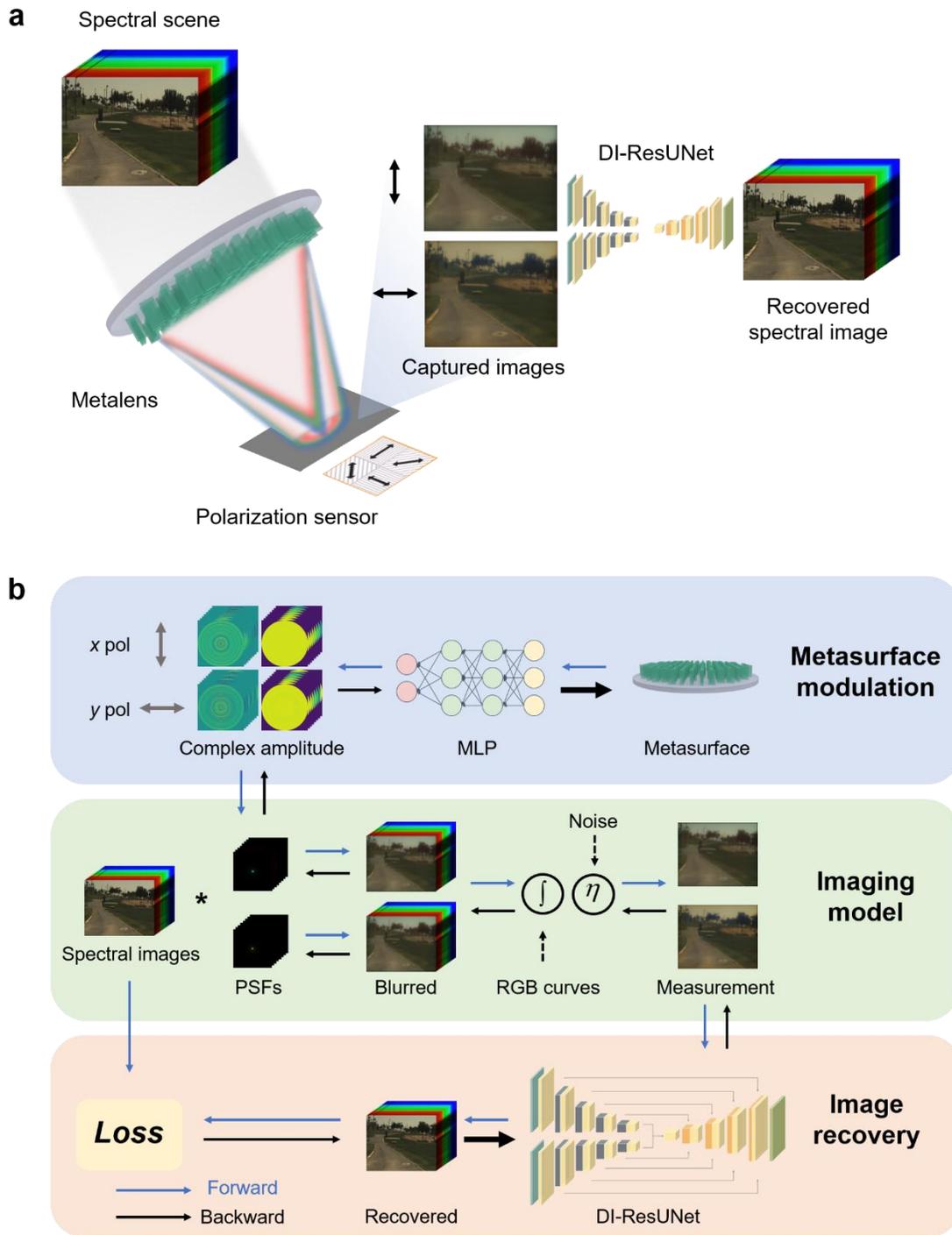

**Fig. 1 Schematic diagram and flowchart of proposed spectral imaging system based on polarization multiplexing metalens.**

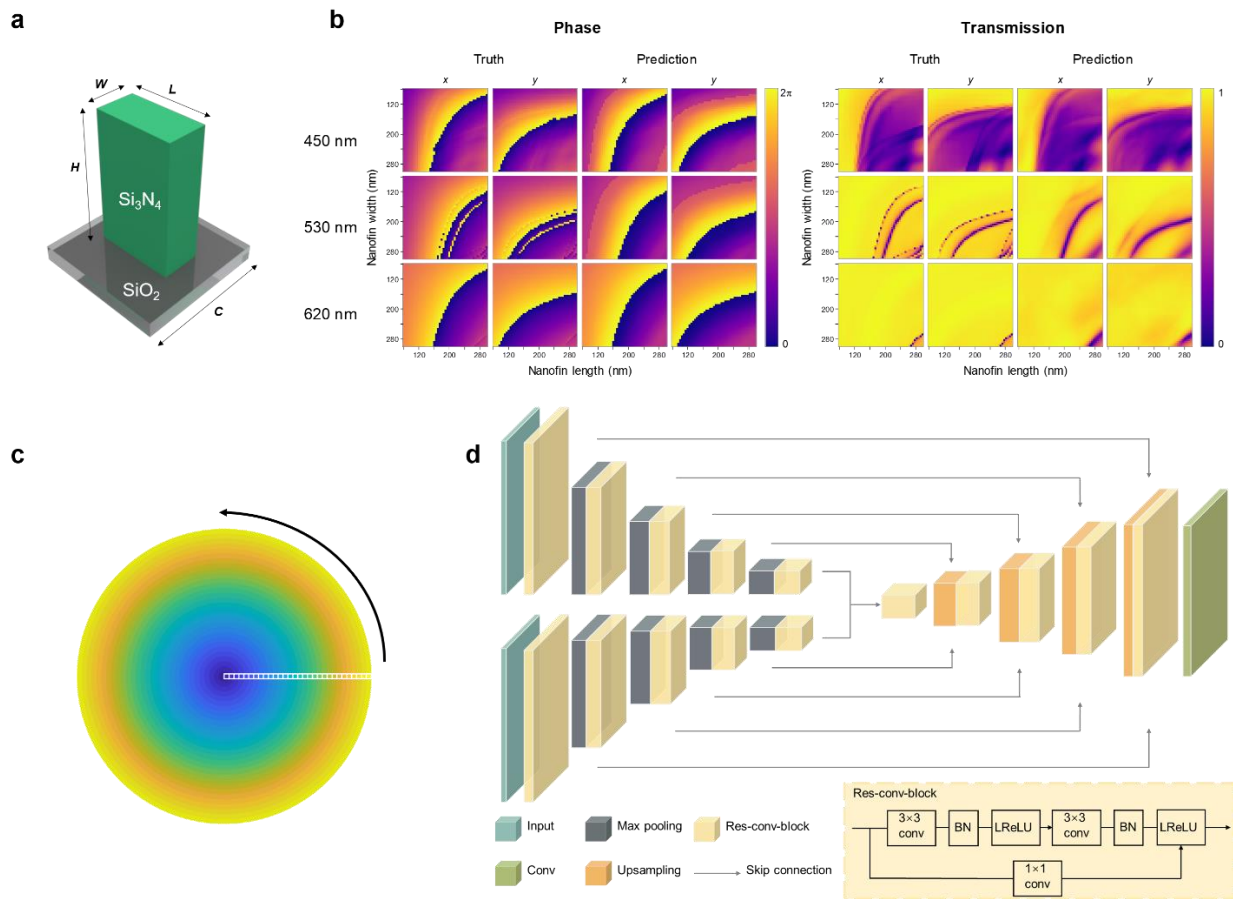

**Fig. 2 Detailed design across various stages of the system. a** Schematic of the unit structure containing a silicon nitride nanofin on a silica substrate. The periodicity C is 350nm, the height of the nanofin H is 800nm, and the length L and width W of the nanofin are sampled evenly every 5nm within the range of 80-300nm. **b** Effectiveness of utilizing MLP for fitting the mapping from unit structure to optical response. 'Truth' and 'Prediction' respectively denote the results of FDTD simulation and the MLP fit. The horizontal and vertical axes in each plot correspond to the length and width of the nanofin. **c** Schematic diagram of rotationally symmetric metalens. **d** Architectural overview of the DI-ResUNet framework, which takes two blurred RGB images as inputs and yields a clear spectral image as the output.

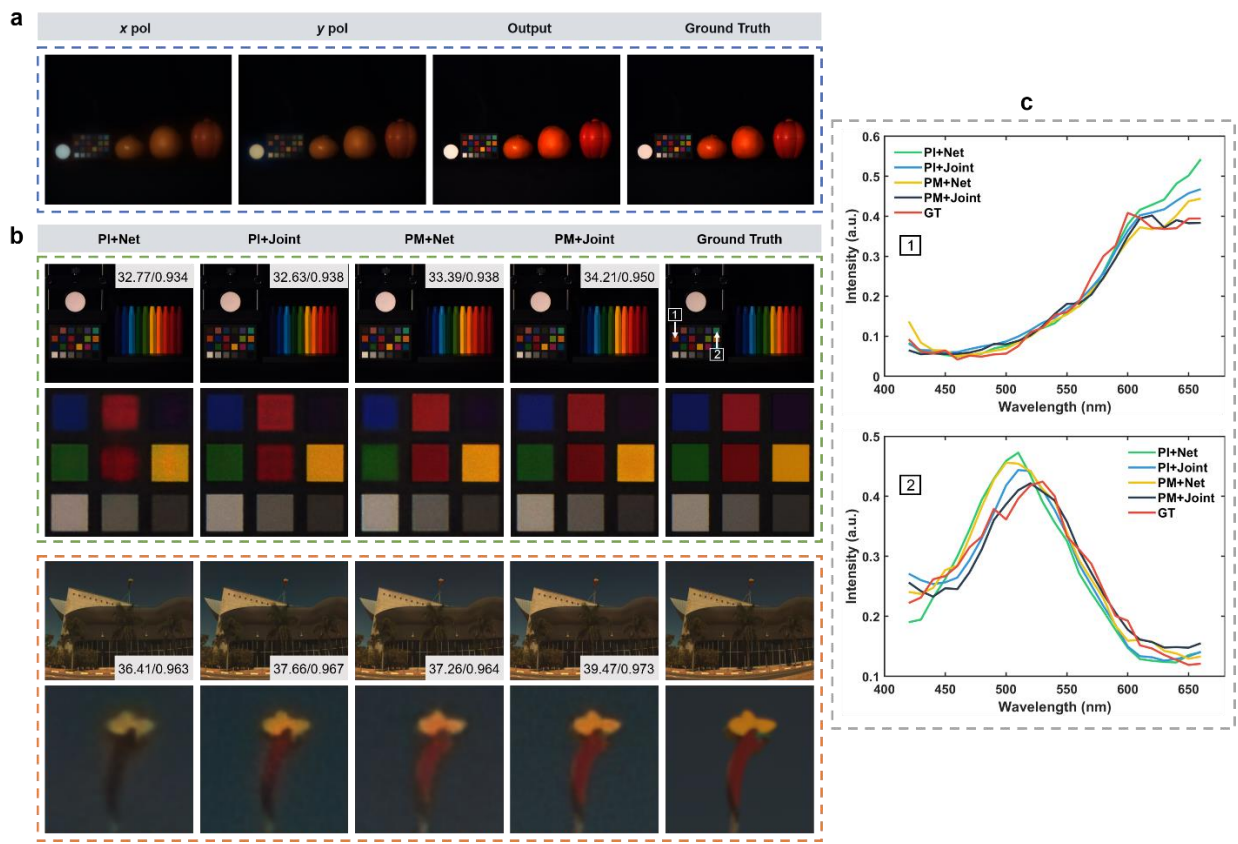

**Fig. 3 Simulated results. a** Simulated RGB measurements captured under x and y polarization states, along with the reconstructed spectral image and corresponding ground truth. The entire 3D spectral data is synthesized into an RGB image for convenience of display in this paper. **b** Visual quality comparison of four methods in simulation. The inset values indicate the PSNR (dB) and SSIM. The images below are magnified views of specific areas. **c** The two panels depict the recovered spectra of points at points 1 and 2 from the color chart of **b** using different methods, respectively.

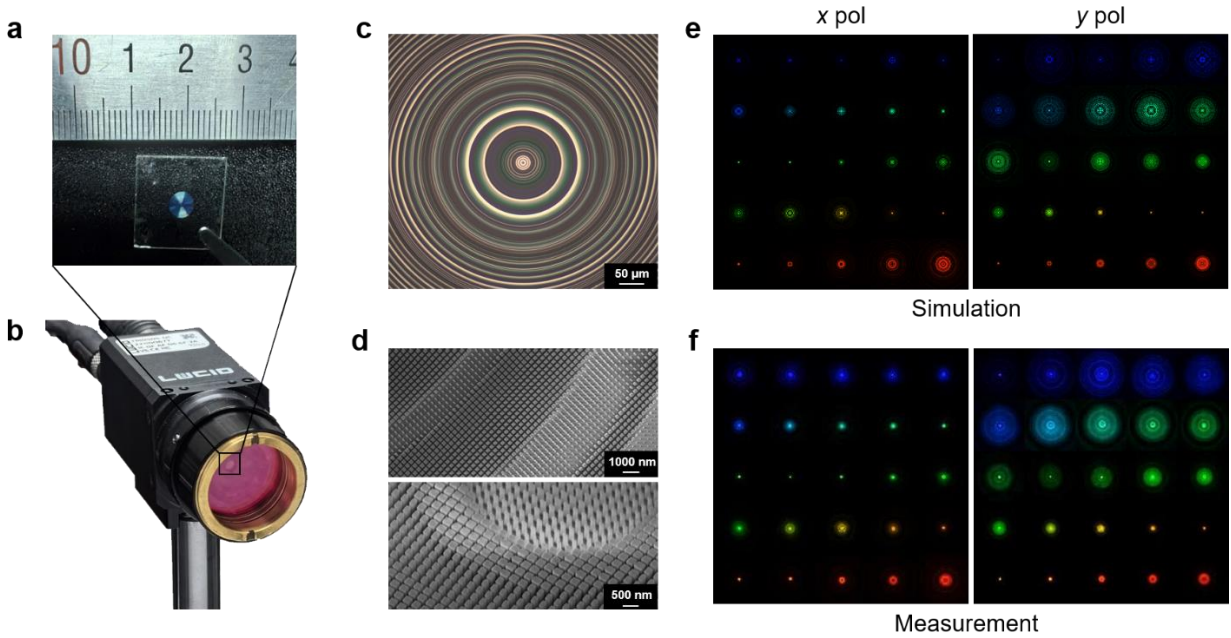

**Fig. 4 Fabricated metalens and PSF measurements. a** Photograph of the fabricated 2.8-mm-diameter metalens. **b** Experimental prototype. **c** Optical microscopy image of the metalens. **d** The scanning electron microscopy (SEM) images show the top view and oblique view of the Si$_3$N$_4$ nanopillars. **e** Simulated PSFs optimized for *x* and *y* polarization channels. **f** Experimentally measured PSFs for *x* and *y* polarization channels.

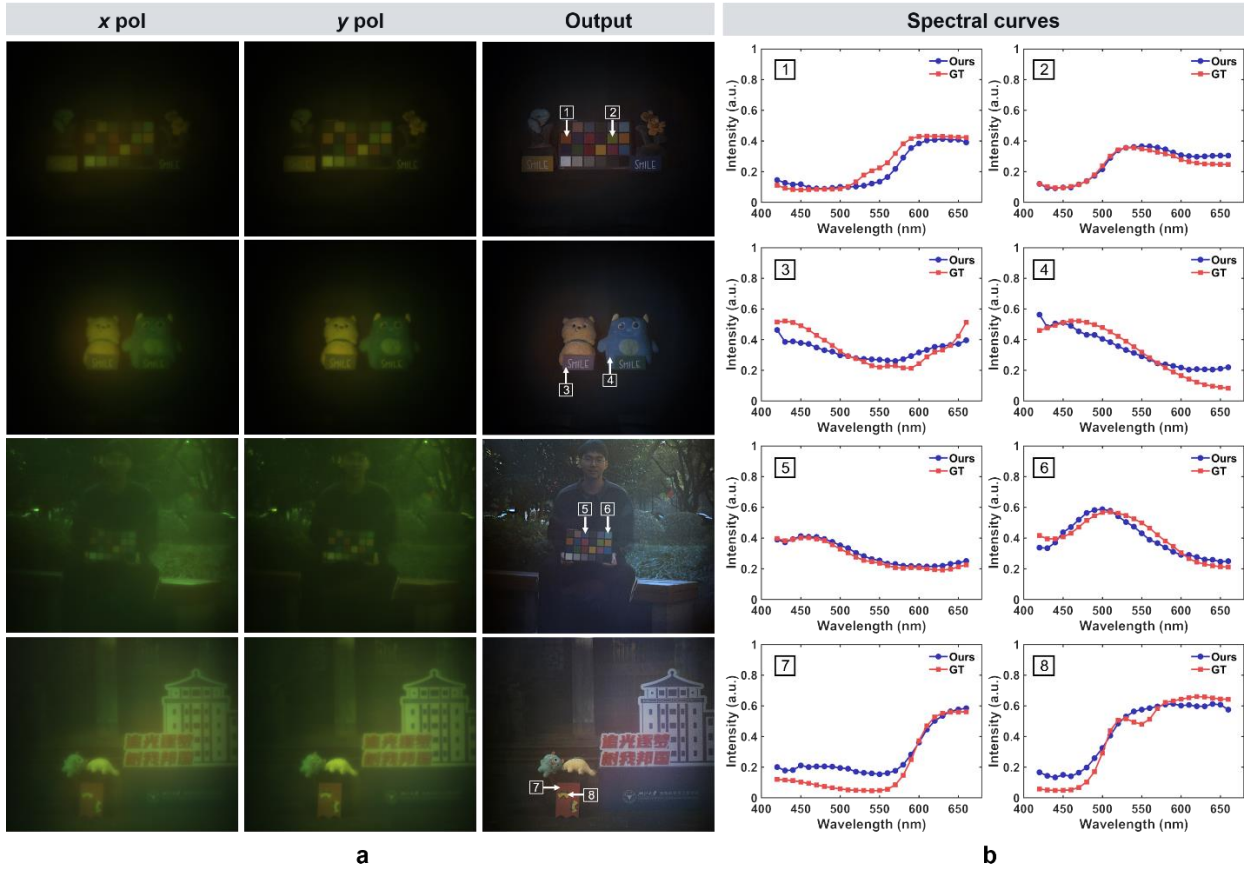

**Fig. 5 Experimental Capture and Reconstruction Results. a** Captured images under *x* and *y* polarization states, accompanied by the corresponding reconstructed spectral images and their detailed insets. **b** Comparison of the spectral reconstruction results for the points marked in **a** with the ground truth measured by a commercial spectrometer.

**Table 1. The numerical results of spectral image recovery on the ICVL and KAIST datasets using four methods. Bold text indicates the highest accuracy.**

| Dataset | ICVL | | | KAIST | | |
|---|---|---|---|---|---|---|
| method | PSNR(dB) | SSIM | RMSE | PSNR(dB) | SSIM | RMSE |
| **PI+Net** | 35.86 | 0.950 | 0.019 | 34.89 | 0.947 | 0.021 |
| **PI+Joint** | 37.20 | 0.962 | 0.016 | 35.90 | 0.950 | 0.019 |
| **PM+Net** | 36.79 | 0.952 | 0.016 | 35.87 | 0.950 | 0.018 |
| **PM+Joint** | **39.00** | **0.966** | **0.012** | **36.98** | **0.962** | **0.017** |